\documentclass[preprint,pre]{revtex4}

\usepackage{graphicx}
\usepackage{dcolumn}
\usepackage{bm}
\usepackage{color}
\usepackage{amssymb}
\usepackage{amsmath}

\begin{document}

\title{Dynamics in Soft-Matter and Biology Studied by Coherent Scattering Probes}

\author{Maikel~C.~Rheinst\"adter}\email{rheinstadter@mcmaster.ca}

\affiliation{Department of Physics and Astronomy, McMaster
University, Hamilton ON, L8S 4M1, Canada, and Canadian Neutron Beam
Centre, Chalk River Laboratories, Chalk River ON, K0J 1J0, Canada}

\date{\today}

\begin{abstract}
Neutrons and x-rays are coherent probes, and their coherent
properties are used in scattering experiments. Only coherent
scattering probes can elucidate collective molecular motions. While
phonons in crystals were studied for half a century now, the study
of collective molecular motions in soft-matter and biology is a
rather new but upcoming field. Collective dynamics often determine
material properties and interactions, and are crucial to establish
dynamics-function relations. We review properties of neutrons and
x-rays and derive the origin of coherent and incoherent scattering.
Taking molecular motions in membranes and proteins as example, the
difference between coherent and incoherent dynamics is discussed,
and how local and collective motions can be accessed in x-ray and
neutron scattering experiments. Matching of coherent properties of
the scattering probe may become important in soft-matter and biology
because of (1) the missing long ranged order and (2) the large
length scales involved. It is likely to be important in systems,
where fluctuating nanoscale domains strongly determine material
properties. Inelastic scattering can provide very local structural
information in disordered systems. Inelastic neutron scattering
experiments point to a coexistence of short-lived nanoscale gel and
fluid domains in phospholipid bilayers in the range of the gel-fluid
phase transition, which may be responsible for critical behavior and
determine elastic properties.
%
%
%

\end{abstract}


\maketitle

\section{Introduction}
Neutron and x-ray scattering techniques were very successfully
applied to investigate structure and dynamics in crystalline
systems. The challenge we face is how to apply these powerful
techniques to soft-matter and biology, i.e., systems with a high
degree of static and dynamic disorder. Even though for instance
biological membranes were studied for decades, very few biologically
relevant processes were revealed on a molecular level. The reason is
the combination of very small nanometer length scales and very fast
dynamics of pico- and nanoseconds, which poses particular
experimental challenges.
Neutron and x-ray scattering is an ideal microscope to study
structure and dynamics in these systems, because they give access to
the relevant length and time scales. Neutrons and x-rays are
coherent probes, and their coherent properties are used in
scattering experiments.
\begin{figure}[]
\centering
\includegraphics[width=1.00\columnwidth,angle=0]{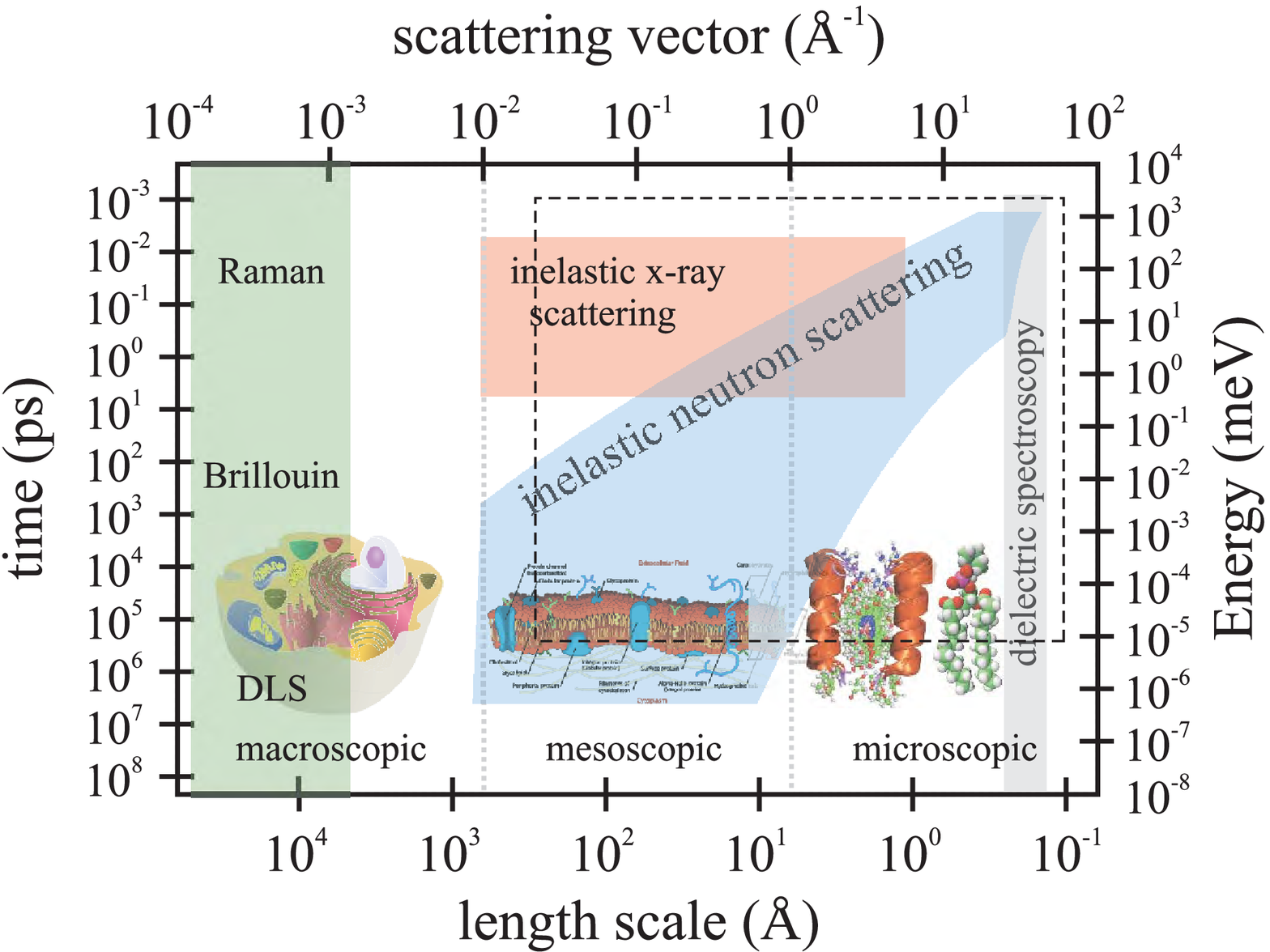}
\caption[]{(Color online). Accessible length and time scales, and
corresponding energy and momentum transfer, for some spectroscopic
techniques covering microscopic to macroscopic dynamics. Light
scattering techniques include Raman, Brillouin, and Dynamic Light
Scattering (DLS). Inelastic x-ray and neutron scattering access
dynamics on Angstrom and nanometer length scales. Dielectric
spectroscopy probes the length scale of an elementary molecular
electric dipole, which can be estimated by the bond length of a C--O
bond (about 140~picometer). The area marked by the dashed box is the
dynamical range accessible by computer simulations.}
\label{spectroscopy}
\end{figure}

Often, the structure of soft-matter and biomolecular systems is
relatively well known and the corresponding experimental techniques,
such as x-ray crystallography, Nuclear Magnetic Resonance (NMR), and
also Atomic Force Microscopy (AFM), developed standard techniques in
many disciplines. Dynamical properties are often less well
understood, but are important for many fundamental properties such
as elasticity and interaction forces. They may determine or strongly
affect certain functional aspects, such as diffusion and transport,
and be relevant for protein function. In particular collective
molecular motions may play a key role to establish dynamics-function
relations in complex soft-matter and biological systems. In this
article, we review properties of neutrons and x-rays and the origin
of coherent and incoherent scattering. The difference between
coherent and incoherent dynamics is discussed, and how local and
collective motions can be accessed in x-ray and neutron scattering
experiments. Matching of coherent properties of probe and sample may
become important in soft-matter and biology because of (1) the
missing long ranged order and (2) the large length scales involved.
Short-lived nanoscale domains in lipid membranes are believed to
strongly determine material properties but their existence is very
difficult to prove. We discuss how inelastic scattering experiments
can provide very local structural information.

Biomolecular and soft-matter materials can be considered as
multi-scale materials because relevant dynamics occur on a large
range of length and time scales. Experimentally, to address the
multi-scale character, different techniques and probes must be
applied. Figure~\ref{spectroscopy} depicts the length and time
scales accessible by inelastic x-ray, inelastic neutron, dynamical
light scattering (DLS), Brillouin and Raman scattering, and
dielectric spectroscopy. By combining different techniques,
macroscopic down to microscopic dynamics can be investigated. The
motions cover slow conformational changes in the millisecond to
microsecond time range. Nanosecond motions include side chain
rotations and backbone torsional reorientations. Rotations of small
side-chains and local vibrational modes motions occur in the fast
picosecond time range. The relevant length scale for dielectric
spectroscopy is in the order of an elementary molecular electric
dipole, which can be estimated by the bond length of a C--O bond
(about 140~picometer), and frequencies from kiloHertz to Gigahertz
can be measured. Because of the large wavelength of the probe
($\lambda_{green}\approx$ 510~nm, $\lambda_{red}\approx$ 632~nm),
light scattering techniques are limited to small momentum transfers
of about $10^{-4}$~\AA$^{-1}$ to $10^{-3}$~\AA$^{-1}$, corresponding
to a length scale of about 100~nanometers. Inelastic neutron and
x-ray scattering access length scales from smaller than Angstrom to
more than 100~nm and time scales from picoseconds to almost one
microsecond. While inelastic x-ray scattering is the perfect tool to
measure fast dynamics at larger distances (small q-values),
inelastic neutron scattering can more easily access slow dynamics at
smaller length scales, as discussed in more detail in
Section~\ref{probes}. Molecular Dynamics (MD) simulations are an
invaluable tool to develop models for molecular structure and
dynamics in membranes and proteins. Because of increasing computer
power and optimized algorithms, large system sizes and long
simulation times, and also more and more complex systems can be
addressed \cite{Smith:1991, Hayward:2002, Tarek:2001, Tarek:2002,
Wood:2007, Meinhold:2007}. The dashed rectangle in
Figure~\ref{spectroscopy} marks the dynamical range accessed by
computer simulations. The elementary time scale in simulations is in
the order of femtoseconds.

Motions in proteins and biological membranes occur on various length
and time scales \cite{Frauenfelder:1991,Fenimore:2004}. The
functional behavior of membrane proteins is likely to depend on the
lipid bilayer composition and physical properties, such as
hydrophobic thickness and elastic moduli. Dynamics in complex
membranes involve interactions between the different constituents,
such as lipids, cholesterol, peptides and proteins. How the variety
of inter- and intra protein motions, occuring over different time
and length scales, interact to result in a functioning biological
system remains an open field for those working at the interface of
physics and biology.
Collective molecular motions in membranes and proteins are
attracting increasing attention
\cite{Meinhold:2007,Liu:2008,RheinstadterPRL:2009}. The reason is
that they may be responsible for certain functionalities, such as
transport of small molecules, pore opening and membrane fusion
processes. New developments and improvements in scattering
instrumentation, sample preparation and environments and,
eventually, the more and more powerful sources make it possible to
access the usually very weak coherently scattered signals.
Collective motions of functional groups may drive protein function.
In a cellular context they may contribute to the understanding of
macromolecular function because they can lead to an effective
coupling and communication between the different constituents. In
contrast to other spectroscopic techniques, such as dielectric
spectroscopy, inelastic x-ray and neutron scattering give a wave
vector resolved access to molecular dynamics. Excitation frequencies
and relaxation rates are measured at different internal length
scales of the system. A typical dynamical scattering experiment
measures (q,$\omega$) pairs, it delivers a frequency together with a
corresponding length scale and possibly also corresponding
direction, such as parallel or perpendicular to the protein axis.
This additional information is very important to assign the measured
frequencies to certain molecules or molecular components.


\section{Local and Collective Molecular Motions}

\begin{figure}[]
\centering
\includegraphics[width=1.00\columnwidth,angle=0]{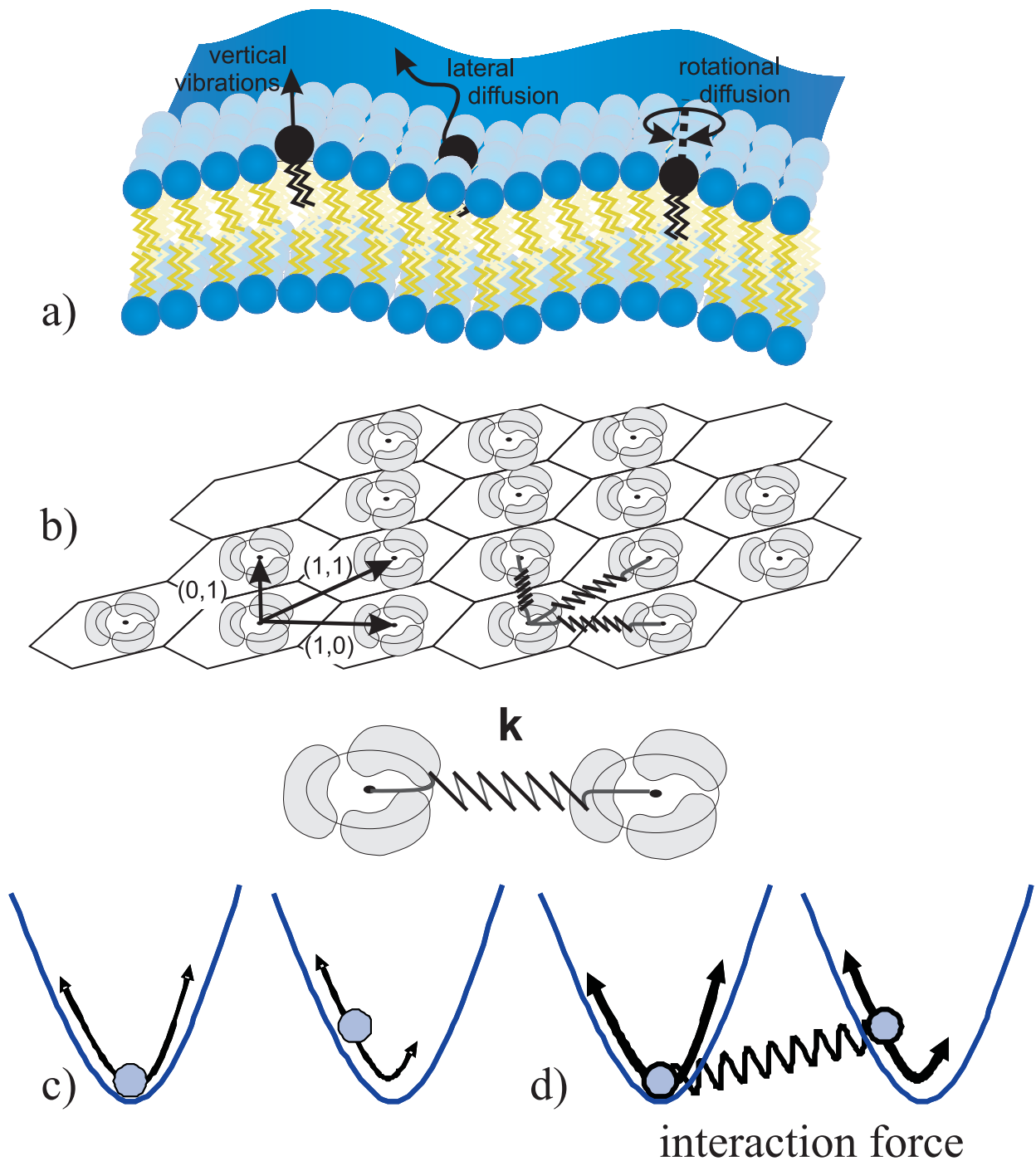}
\caption[]{(Color online). (a) Some elementary dynamical modes in
lipid bilayers. Local modes include diffusion and vibrations,
rotations, and librations (hindered rotations) of single lipid
molecules. (b) Coupling of membrane embedded proteins in a hexagonal
arrangement, taking bacteriorhodopsin in purple membrane as an
example. The interaction between protein trimers is depicted as
springs with effective spring constant k. (c) Incoherent dynamics
stems from the motion of molecules or functional groups in local
energy potentials. (d) Coherent dynamics involve interactions of
particles and probe the interaction force.}
\label{membranedynamics.eps}
\end{figure}

Atomic and molecular motions can be classified as local,
self-correlated, and collective, pair-correlated dynamics. Local
dynamics is the motion of molecules or functional groups in local
energy potentials. The force and time constants involved are
determined by the local friction and restoring forces. This type of
dynamics is called incoherent because the particles move
independently in their local environments and examples are
vibration, rotation, libration (hindered rotation) and diffusion of
individual lipid molecules.

Coupled, collective molecular motions need an interaction between
neighboring particles or functional groups (pictured as a spring).
Coupled particles basically behave like coupled pendulums. This type
of dynamics is called coherent. Collective molecular motions
determine for instance elasticity of membranes, interactions between
membrane embedded proteins, and may impact on transport processes.
In biology, any dynamics will most likely show a mixed behavior of
particles moving in local potentials but with a more or less
pronounced coherent (interacting) character.
Figure~\ref{membranedynamics.eps} exemplary depicts some of the
local and collective modes in a lipid bilayer. Rotational and
lateral diffusion, vibrations and rotations of the single lipid
molecules can be investigated by, e.g., incoherent inelastic neutron
scattering, nuclear magnetic resonance or dielectric spectroscopy
(Figure~\ref{membranedynamics.eps} (a) and (c)). Only coherent
probes, such as coherent inelastic neutron scattering or inelastic
x-ray scattering, can elucidate collective molecular motions, as
will be discussed below. Collective molecular motions can for
instance be depicted by a system of coupled membrane embedded
proteins in Figure ~\ref{membranedynamics.eps} (b). The
corresponding interaction forces (Figure ~\ref{membranedynamics.eps}
(d)) can be determined in scattering experiments.

\subsection*{Coherent and Incoherent Scattering}  The peculiarity of
neutrons is that they can scatter incoherently or coherently, and
therefore give access to local or collective dynamics. If a single
atom of an element is hit by a photon or neutron represented by a
planar wave, $\psi_{inc}=e^{ikz}$, the scattering can be analyzed in
terms of an emerging spherical S-wave , $\psi_{sc}=-b/r \cdot
e^{ikr}$. The amplitude of the scattered spherical wave, b, is
defined as the scattering length of the atom. The scattering length
of an element is different for x-rays and neutrons. Because x-rays
are electromagnetic waves and scattered by the electrons in the
electron shell, the scattering length depends on the number of
electrons Z, and is proportional to the element number, b$\propto$Z
\cite{AlsNielsen:2001}.

The case is different for the interaction of neutrons with matter.
Neutrons are microscopic particles. They carry a mass of m=1.675
10$^{-27}$ kg and move with velocities between about 500 m/s (cold
neutrons) to 2,200 m/s (thermal neutrons). Neutrons carry no charge,
but a magnetic moment, a spin $\frac{1}{2}$ \cite{Squires:1978}. In
contrast to x-rays, neutrons are scattered by the atomic nucleus.
The scattering length b depends on the spin of the nucleus-neutron
system. If the spin of the nucleus is I, then every nucleus with non
zero spin has two values of b, namely for I$+\frac{1}{2}$ and
I$-\frac{1}{2}$, depending on the orientation of neutron and nucleus
spin. The scattering length is not simply a monotonic function of
the atomic number, as it is for x-rays, but depends on the spin
configurations of the nuclei. In addition, different isotopes of the
same element have different scattering lengths. Isotopes have nuclei
with the same number of protons (the same atomic number) but
different numbers of neutrons and also different nuclear spins I. So
while for x-rays, all atoms of the same element look the same, they
may look different for neutrons because of (1) different
orientations of the nuclear spin, and (2) different isotopes
(nuclides). The most pronounced difference between the x-ray and the
neutron probe is therefore that x-ray scattering is always coherent,
while neutron scattering may contain contributions from coherent and
incoherent scattering, as will be discussed in the following.

The coherent and incoherent neutron cross sections, b$_{coh}$ and
b$_{inc}$, of an element can be illustrated by two extreme cases. If
all the nuclei in a sample have different b's, there is no
interference between the waves scattered by the different atoms. The
incoherent scattering therefore depends only on correlation between
the positions of the same nucleus at different times. Incoherent
scattering therefore probes the local atomic or molecular
environment. If all scattering lengths are the same, i.e., all
nuclei are identical for the neutron probe, the coherent scattering
still depends on the correlation between the positions of the same
nucleus at different times, but also on the correlation between the
positions of different nuclei at different times. It therefore gives
interference effects and allows to measure interaction forces. In
general, every element has coherent and incoherent scattering
lengths. Table~\ref{scatteringlength} lists the scattering lengths
for selected elements, which are common in biological material.

\begin{table}
\centering
\begin{tabular}{|c|c|c|c|c|c|}
  \hline
    Nuclide & b$_{coh}$ (fm) &   b$_{inc}$ (fm) &  Nuclide & b$_{coh}$ (fm) &  b$_{inc}$
    (fm) \\\hline
$^1$H &  -3.7406 & 25.274 & $^{12}$C & 6.6511 & 0\\
$^2$H & 6.671 &  4.04  &  $^{13}$C & 6.19  &  -0.52\\
$^3$H & 4.792 &  -1.04  & $^{16}$O & 5.803 &  0\\
$^{14}$N & 9.37  &  2.0 & $^{17}$O & 5.78  &  0.18\\
$^{15}$N & 6.44 &   -0.02 &  $^{18}$O & 5.84 &   0\\
  \hline
\end{tabular}
\caption[]{Coherent and incoherent scattering lengths for selected
elements as provided by the NIST Center for Neutron Research
(http://www.ncnr.nist.gov/resources/n-lengths/).}\label{scatteringlength}
\end{table}

b$_{coh}$ is defined as the average scattering length, $\bar{b}$.
The terms $\bar{b}$ and $\bar{b^2}$ can be used to define the
coherent and incoherent scattering cross sections,  $\sigma_{coh}$
and $\sigma_{inc}$ by $\sigma_{coh}=4\pi\bar{b}^2$ and
$\sigma_{inc}=4\pi\left( \bar{b^2}-\bar{b}^2\right)$, which are more
commonly used. The unit of the cross section is 1 barn (1 b),
equivalent to an area of 10$^{-28}$ m$^2$. The coherent scattering
is then the scattering of a system where all scattering lengths are
equal to $\bar{b}$. The incoherent scattering stems from the random
distribution of deviations of the scattering length from the mean
value $\bar{b}$. Table~\ref{crosssections} lists values for the
scattering cross sections for selected elements. Most noticeable are
the very large incoherent cross section for hydrogen, $^1$H, and the
relatively large coherent cross section for deuterium, $^2$H. This
is important for selective deuteration and labeling, as will be
discussed in one of the next paragraphs.

\begin{table}
\centering
\begin{tabular}{|c|c|c|c|c|c|}
  \hline
Nuclide & $\sigma_{coh}$ (b) &  $\sigma_{inc}$ (b) &   Nuclide &
$\sigma_{coh}$ (b) & $\sigma_{inc}$ (b)\\\hline

$^1$H & 1.7583 & 80.27 & $^{12}$C & 5.559 & 0\\
$^2$H & 5.592 & 2.05 & $^{13}$C & 4.81 & 0.034 \\
$^3$H & 2.89 & 0.14 & $^{16}$O & 4.232 & 0 \\
$^{14}$N & 11.03 & 0.5 & $^{17}$O & 4.2 &
0.004\\
$^{15}$N & 5.21 & 0.00005 & $^{18}$O & 4.29 & 0\\
  \hline
\end{tabular}
\caption[]{Coherent and incoherent scattering cross sections, as
provided by the NIST Center for Neutron Research
(http://www.ncnr.nist.gov/resources/n-lengths/).}\label{crosssections}
\end{table}

\subsection*{Scattering Functions}
To more accurately define coherent and incoherent scattering, the
corresponding scattering functions shall be introduced briefly. The
differential cross section of the quasi-elastic scattering of
neutrons in a solid angle $\mbox{d}\Omega$ with an energy transfer
$\hbar \omega$ can be expressed as \cite{Lovesey:1984}
\begin{equation}
\frac{\mbox{d} \sigma^2}{\mbox{d} \Omega \mbox{d} \omega} \propto
\sum_{n} \left( b_{inc}^n \right)^2 S_{inc}^{n}(q,\omega) + \sum_{n,
m} b_{coh}^n b_{coh}^m S_{coh}^{n m}(q,\omega), \label{crosssection}
\end{equation}
where $n, m$ are atom-type indices, while $b_{inc}^n$ [$b_{coh}^n$]
and $S_{inc}^n(q,\omega)$ [$S_{coh}^{n m}(q,\omega)$] are,
respectively, the incoherent [coherent] scattering length and
dynamic structure factor. $S(q,\omega)$ is directly measured in
inelastic x-ray and neutron scattering experiments.

$S_{inc}^n(q,\omega)$ is obtained from the Fourier transform of the
self intermediate scattering function
\begin{equation}
I_{self}^n(q,t) = \frac{1}{N_n} \left\langle \sum_{j=1}^{N_n} e^{i
\vec{q} \cdot [ \vec{r}\,_j^n(0) -
  \vec{r}\,_j^n(t)]} \right\rangle ,
\label{inc}
\end{equation}
where the summation goes over all $N_n$ atoms of type $n$ in the
system. The coherent part, $S_{coh}^{n m}(q,\omega)$, involves
correlations between different particles and can be obtained from
the Fourier transform of the intermediate scattering function
\cite{Squires:1978}
\begin{equation}
I^{n,m}(q,t) = \frac{1}{N_n N_m} \left\langle
\sum_{j,j'=1}^{N_n,N_m} e^{i \vec{q} \cdot [ \vec{r}\,_j^n(0) -
  \vec{r}\,_{j'}^m(t)]} \right\rangle .
\label{coh}
\end{equation}

The coherent scattering function $S_{coh}^{nm}(q,
\omega)=\frac{1}{2\pi\hbar}\int{I^{nm}(q,t)e^{i\omega t} dt}$
describes the time evolution of positional correlations between
different molecules or atoms in a system. In the incoherent
scattering function, $S_{inc}^{n}(q,
\omega)=\frac{1}{2\pi\hbar}\int{I_{self}^{n}(q,t)e^{i\omega t} dt}$,
positional correlations of the same particle at different times are
considered, only. While $S_{coh}^{nm}(q, \omega)$ describes
pair-correlated dynamics, the incoherent scattering, $S_{inc}^{n}(q,
\omega)$, describes self-correlated motions. Interactions between
proteins in a membrane are an example for coherent dynamics and the
corresponding (coherent) scattering experiments measure the
interaction force between the proteins, as depicted in
Figure~\ref{membranedynamics.eps} (d). Diffusion of a single lipid
molecule for instance is observed as incoherent scattering. An
(incoherent) scattering experiment probes the local energy
potential, as sketched in Figure~\ref{membranedynamics.eps} (c). But
how can neutron scattering experiments be tuned to measure coherent
respective incoherent dynamics and what is the difference between
x-ray and neutron spectra?

\subsection*{Selective Deuteration and Labeling Techniques}  The total
scattering of a sample in neutron scattering experiments will
contain coherent and incoherent scattering contributions. The
fraction of coherent and incoherent scattering depends on the atomic
composition and the respective scattering lengths. Substitution of
certain elements in a compound by their isotopes may increase
contributions of certain molecules or functional groups to the
coherent or incoherent scattering contribution.

The incoherent cross section of hydrogen atoms is about 40 times
larger than that of deuterium (see Table~\ref{crosssections}), and
of all other atoms present in biological macromolecules.
Consequently, hydrogen atoms dominate the incoherent scattering
signal of biological samples. The hydrogen atoms reflect the
movements of larger groups to which they are attached, such as amino
acid side chains. Deuteration, i.e., the substitution of protons by
deuterium ($^2$H), is often used to increase or suppress the
incoherent scattering contribution of certain functional groups to
the total scattering.

\begin{figure}[]
\centering
\includegraphics[width=1.00\columnwidth,angle=0]{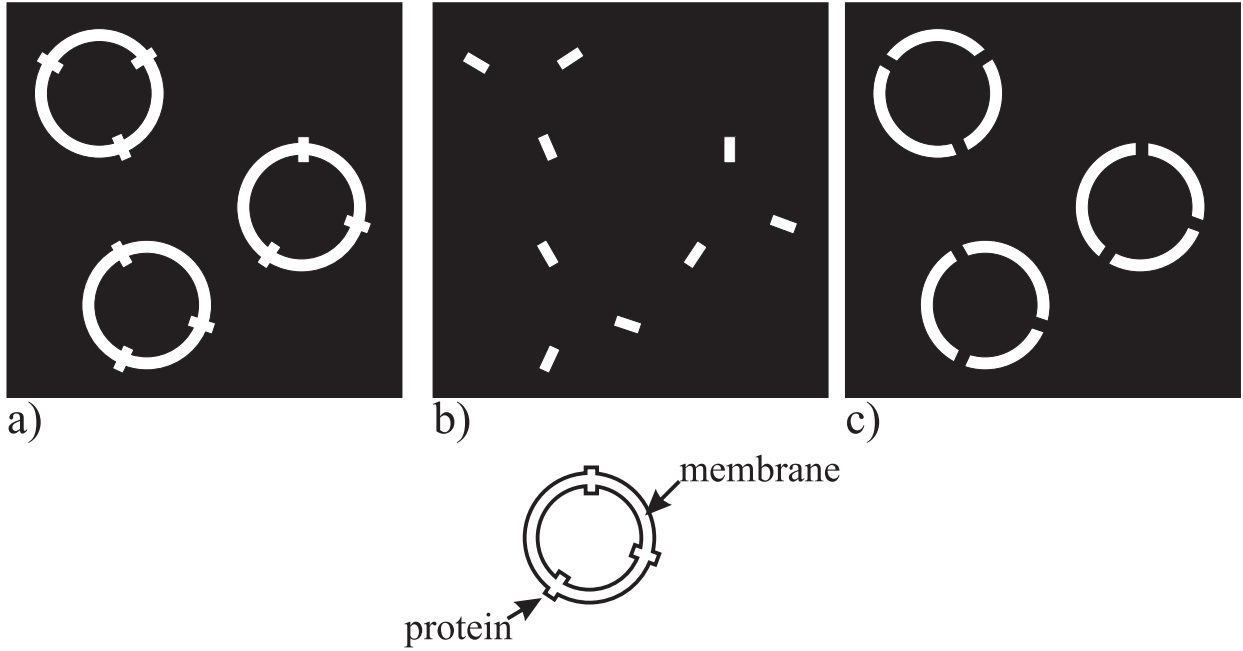}
\caption[]{Selective deuteration can be used to highlight molecules
or functional groups in neutron scattering experiments. Depicted are
vesicles with embedded proteins in a solvent: (a) membrane and
embedded proteins are visible in incoherent scattering experiments
for neutrons when the solvent is deuterated. (b) When solvent and
membrane are deuterated, local dynamics of the embedded proteins can
be investigated. (c) Deuterated proteins allow investigating
membrane dynamics.} \label{selectvedeuteration}
\end{figure}

While in protonated samples the incoherent scattering is usually
dominant and the time self-correlation function of individual
scatterers is accessible, (partial) deuteration emphasizes the
coherent scattering and gives access to collective motions by
probing the pair correlation function. In a membrane sample with
protonated proteins, the experiments would be sensitive to the
diffusive motions of the proteins. Deuteration of the proteins
increases the coherent scattering and allows measuring the
interaction forces between the embedded proteins. The effect of
labeling is sketched in Figure~\ref{selectvedeuteration}. In
Figure~\ref{selectvedeuteration} (a), self-correlated, diffusive
motions of membranes and embedded proteins can be accessed when the
membrane-protein system is labeled, and deuterated solvent is used.
Diffusive dynamics of the proteins is highlighted when solvent and
membrane are deuterated, and hydrogenated proteins are used. The
effect of protein insertion on membrane dynamics can be studied in
Figure~\ref{selectvedeuteration} (c), with deuterated solvent and
proteins. Note that at the same time, the interfaces between
protonated and deuterated areas scatter coherently. The preparation
in Figure~\ref{selectvedeuteration} (b) can therefore be used to
study possible protein-protein interactions.

\subsection*{Excitation Spectra: The Finger Print of Dynamics}
\begin{figure}[]
\centering
\includegraphics[width=1.00\columnwidth,angle=0]{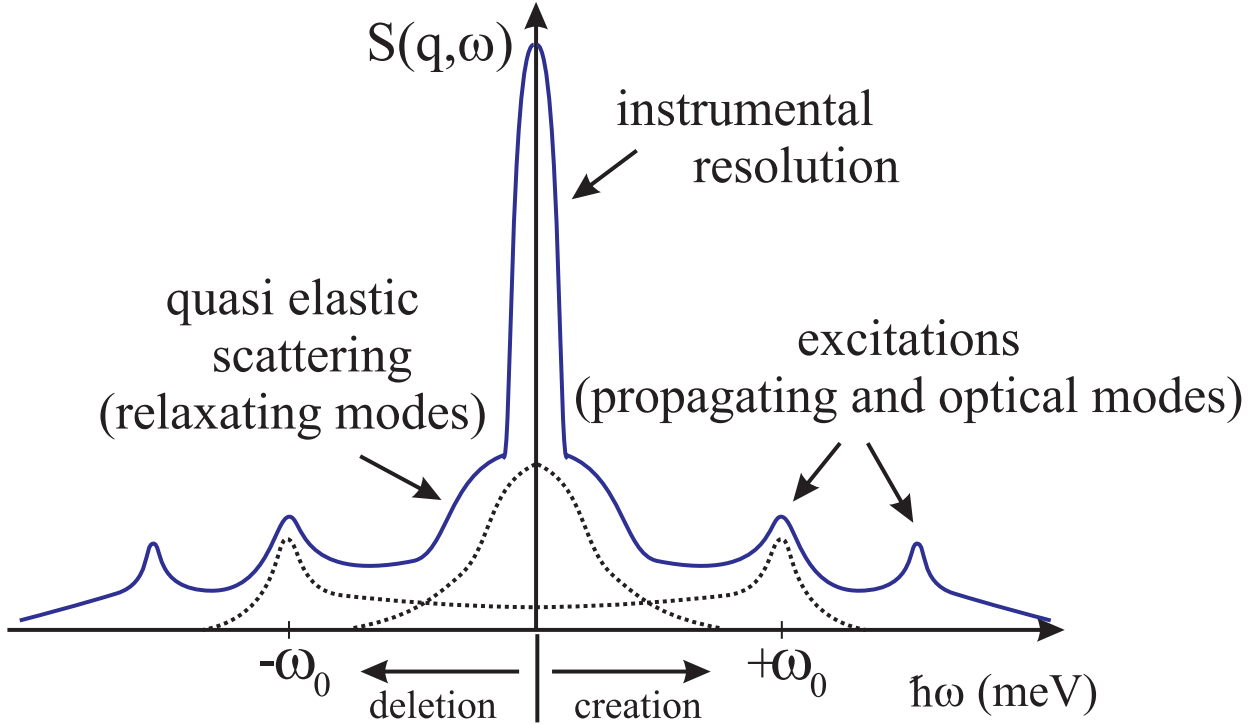}
\caption[]{(Color online). Typical excitation spectrum for x-rays
and neutrons. The signal consists of a sharp central line, the
instrumental resolution, a broad component centered at energy
transfer zero (quasielastic scattering), and pairs of excitations
from deletion and creation of dynamical modes.} \label{excitations}
\end{figure}
While Bragg scattering, i.e., the occurrence of diffraction peaks,
stems from purely coherent scattering, inelastic spectra may contain
contributions from coherent and incoherent scattering. A typical
excitation spectrum $S(q,\omega)$, measured by inelastic x-ray or
neutron scattering at a particular q-vector, is shown in
Figure~\ref{excitations}. The total signal typically consists of a
sharp peak at energy transfer zero, which marks the instrumental
resolution. The resolution is often well described by a Gaussian
peak shape. Quasielastic scattering is observed as a broadening of
the instrumental resolution. In addition, inelastic peaks may be
observed. Excitations always occur as pairs of inelastic peaks at
frequencies of $-\omega_0$ and $+\omega_0$, caused by creation and
deletion of dynamical modes. Starting from the simple models for
local and collective modes in Figure~\ref{membranedynamics.eps} (c)
and (d), assumptions for peak shapes of the corresponding modes can
be developed. Typical local modes, such as diffusion and libration
(hindered rotation), are examples for relaxators, i.e., overdamped
oscillations. The characteristic frequency of a relaxator is
$\omega_0=0$, so that the response is centered on the elastic line.
The width of the response is determined by the relaxation time
$\tau_0$ as $\Delta\hbar\omega=2\pi/\tau_0$. In the time domain a
relaxation is described by an exponential decay $e^{-t/\tau_0}$ ,
the Fourier transform of which is a Lorentzian function. So local,
incoherent dynamics give rise to a quasielastic broadening in the
energy domain and are described by Lorentzian line shapes, centered
at energy transfer zero. Coherent dynamics, probe the elastic
springs between molecules or atoms. They can be considered as
particles bound by an isotropic harmonic force and damped through
hybridization with a phonon bath. Those propagating (acoustical) or
oscillating (optical) modes have well defined eigenfrequencies and
lead to inelastic excitations at energy values $\pm\hbar\omega_0$.
They are described by a damped harmonic oscillator model (dho). The
width of the inelastic peak is related to the life time of the
excitation. A strict derivation of the peak profiles can be found in
the books by Squires \cite{Squires:1978} and Lovesey
\cite{Lovesey:1984}.

The excitation spectrum looks identical for the x-ray and neutron
probes. The origin of quasielastic and inelastic scattering is
purely coherent for x-rays. In the case of liquids for instance, the
spectrum can be evaluated by a generalized effective eigenmode
theory \cite{Liao:2000} in terms of a heat mode (a collective
diffusion process) and sound modes. For neutrons, the quasielastic
part of the spectrum can have contributions from coherent and
incoherent scattering, but is most likely dominated by the
incoherent scattering of hydrogen atoms. The corresponding dynamical
modes are relaxating local modes, such as diffusion of molecules or
functional groups. The relaxation times can be determined from the
width of a Lorentzian peak shape. Excitations stem from coherent
scattering and involve dynamics between different particles from
interference effects. So by preparing a hydrogenated or deuterated
sample and tuning the spectrometer to quasi- or inelastic
scattering, a neutron scattering experiment can be made sensitive to
local or collective dynamics of certain components to probe local
environments or interactions.

\subsection*{Scattering Probes and Dispersion Relation\label{probes}}
\begin{figure}[]
\centering
\includegraphics[width=1.00\columnwidth,angle=0]{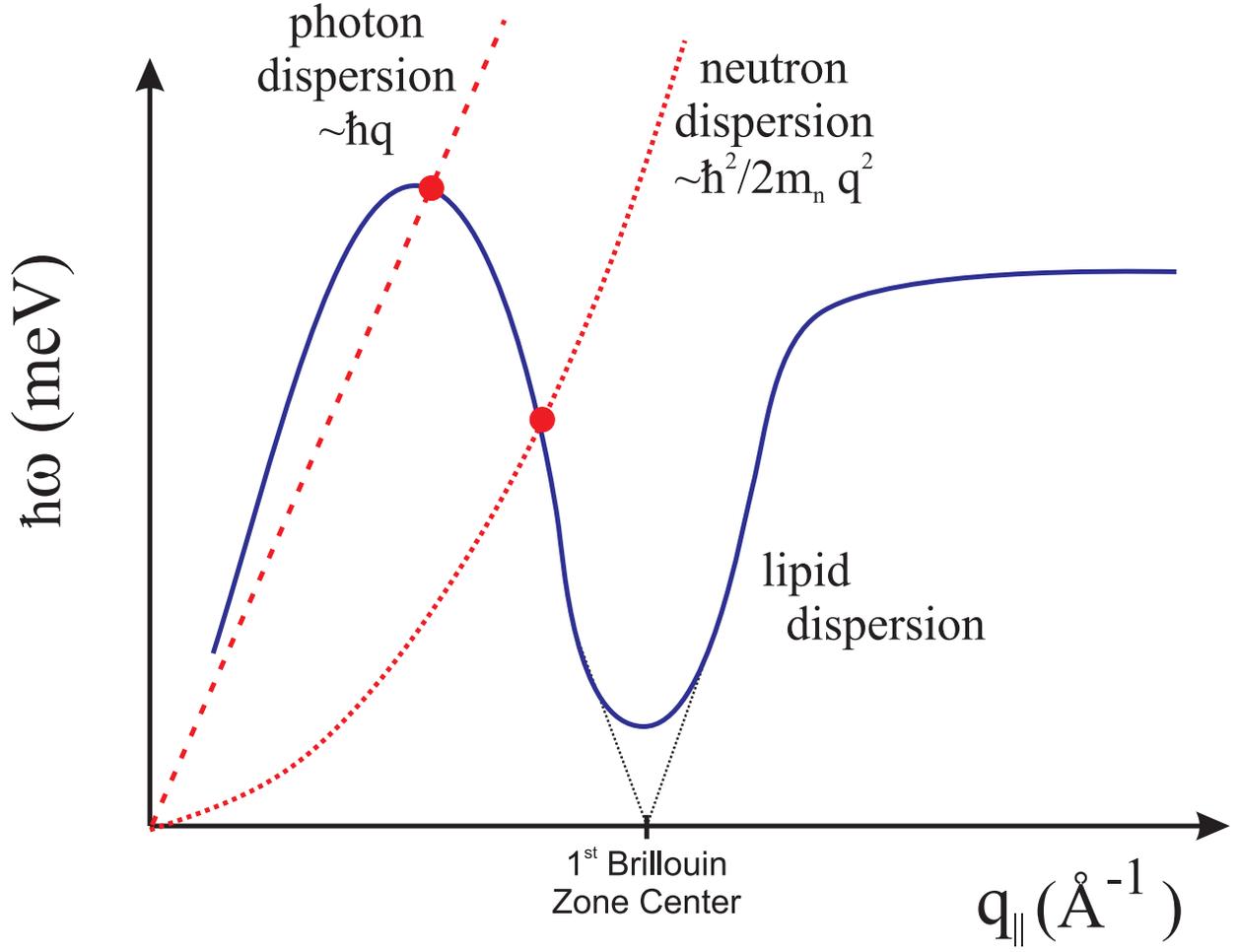}
\caption[]{(Color online). Dispersion relations of a phospholipid
bilayer (adapted from \cite{RheinstaedterPRL:2004}), and the neutron
and x-ray probe. The intersections mark positions where energy and
momentum conservation is fulfilled.} \label{dispersions}
\end{figure}
Collective motions can be modeled by interacting particles, as
sketched in Figures~\ref{membranedynamics.eps} (b) and (d). The
dynamics must be described by a set of coupled pendulums (particles
bound by an isotropic harmonic force). In inelastic scattering
experiments, excitation energies are probed at different internal
length scales and different directions. No trivial q-dependence of
the energies can usually be expected. The energies will depend on
the interaction constant, k, but also on the geometrical arrangement
of the objects. It will depend on the length scale, the q-value, but
critically also on the coupling path, i.e., the direction, which is
probed, as depicted in Figure 2 (b). A typical experiment to measure
collective dynamics records excitation frequencies at different
q-values over a coherent q-range. The corresponding curve
$\hbar\omega(q)$ is called a dispersion relation. It describes the
complex dynamics of a (oriented) system of coupled oscillators.
Technically, it describes the way waves with a certain wavelength,
and certain coherence length, propagate through the system. Note
that only coherent probes are capable to elucidate coherent
dynamics.

Figure~\ref{dispersions} exemplary depicts the dispersion relation
of the collective short wavelength density fluctuations in a
phospholipid bilayer, as adapted from \cite{RheinstaedterPRL:2004}.
$q_{||}$ is the lateral momentum transfer, i.e., the projection of
the scattering vector $Q$ into the plane of the membrane. The
particular shape of the corresponding dispersion relation can
qualitatively be explained. The basic scenario is the following: at
small $q_{||}$, longitudinal sound waves in the plane of the bilayer
are probed and give rise to a linear increase of $\omega\propto
q_{||}$, saturating at some maximum value, before a pronounced
minimum is observed at $q_0\approx$1.4~\AA$^{-1}$, the first maximum
in the static structure factor $S(q_{||})$ (the inter acyl chain
correlation peak). $q_0$ can be interpreted as the quasi- Brillouin
zone of a twodimensional liquid. Collective modes with a wavelength
of the average nearest neighbour distance $2\pi/q_0$ are
energetically favorable. The static and dynamic disorder in the
lipid bilayers finally leads to a minimum at finite energy values
(softmode). The lipid dispersion relation is similar to those in
ideal liquids, as, e.g. liquid argon \cite{Schepper:1983,
Well:1985}, liquid neon \cite{Well_3:1985} or liquid helium
\cite{Glyde:1994}. The interior of a lipid bilayer, the C-atoms or
C–H groups of the lipid acyl chains, behave like a quasi liquid. In
contrast to real liquids, the chain atoms of the lipid molecules are
chemically bound to each other, leading to smaller mobility and
diffusion and, as a consequence, more pronounced excitations. Higher
Brillouin zone centers are usually not well developed in soft-matter
and biology because of the high degree of disorder. Scattering
experiments are therefore usually limited to the 0$^{th}$ Brillouin
zone, only. This is true in particular for inelastic scattering
experiments and probably one of the most fundamental differences to
scattering experiments in crystalline systems. One of the
consequences is that momentum transfer in soft-matter and biology
occurs mainly in the forward direction. The transverse component is
usually very small, in contrast to crystals, where longitudinal and
transverse branches are found.

The dispersion relations $\hbar\omega(q_{||})$ of the neutron and
x-ray scattering probes are also plotted in
Figure~\ref{dispersions}. The dispersion of the photon probe is
proportional to $\hbar q_{||}$ (with photon energies of keV), i.e.,
a very steep straight line as compared to typical thermal excitation
energies in biological materials in the meV range. Because of the
distinct particle character of neutrons, the corresponding
dispersion relation is proportional to $\hbar^2/(2 m_n) q_{||}^2$
(with neutron mass $m_n$). Neutrons carry kinetic energies
comparable to thermal energy (``thermal neutrons''). For soft-matter
and biological studies, which usually involve large distances and
low excitation energies, neutrons are often slowed down by a cold
moderator to produce ``cold (slow) neutrons'' to further enhance the
energy and $q$-resolution. Momentum and energy are conserved during
the scattering process. As a graphical interpretation of the
conservation laws, the dispersion curves of probe and lipids must
intersect, as shown in Figure~\ref{dispersions}. The range of energy
and momentum transfers accessible by the respective neutron or x-ray
instrument depends on mechanical restrictions, such as angular
range, but also on the properties of the scattering probe. X-rays
can easily access faster dynamics at small $q$-values (Brillouin
scattering), but offer lower energy resolution because of the high
incident photon energies. Thermal and cold neutrons offer a high
energy resolution because incident energies are comparable to the
energies of dynamical modes. The parabolic shape of the neutron
dispersion prevents inelastic experiments at small $q$-values and at
the same time high energy transfers. This is called the kinematic
restriction.

\subsection*{Coherence\label{coherence}}
While in most natural systems, waves with only partial coherence are
more common, x-rays and neutrons both are ``coherent'' probes with
coherence lengths significantly longer than common lattice spacings,
and even protein-protein distances in biological membranes. The
coherence of waves in periodic systems (lattices) is crucial to
their dynamics, as interference effects, such as Bragg reflections,
largely determine their propagation. Also in less well ordered
systems, the coherence length of the probe, $\lambda_C$, may play an
important role for the investigation of small structures, comparable
to the size of $\lambda_C$. The coherence properties of x-rays and
neutrons depend on their wave character.

A collimated beam of neutrons or x-rays prepared by Bragg reflection
from a single crystal, or by passage through a pair of phased Fermi
choppers, is never precisely monochromatic. From the wave
properties, the longitudinal coherence length, $\lambda_C$, can be
estimated to be $\lambda_C=\lambda^2/\Delta\lambda$
\cite{Rauch:1993}. A longitudinal coherence length of 100~\AA\ was
reported for thermal neutrons with wavelength of $\lambda$=1.8~\AA
  \cite{Rauch:1996}. For cold neutrons with wavelengths of about
5~\AA\, and a typical monochromaticity by single crystal reflection
of about 5\%, the longitudinal coherence length can be estimated to
be about 500~\AA. Note that the reason for the typically rather low
monochromaticity of neutron beams is not to compromise the
relatively low intensity of the neutron sources, as compared to
x-ray synchrotron sources. Longitudinal coherence for x-rays
reflected from a Si(111) monochromator with a wavelength resolution
of $\Delta\lambda/\lambda\approx 1 \cdot 10^{-4}$ and a wavelength
of $\lambda$=1~\AA\ is in the order of
$\lambda_C^{xray}$=10,000~\AA. The coherence properties of the
scattering probe may play an important role for the investigation of
small structures, such as nanoscale domains, comparable to or
smaller than the size of $\lambda_C$. Structures smaller than
$\lambda_C$ may lead to spatially averaged values for, e.g., peak
positions and widths. As will be shown below, inelastic data may
still show separate excitations for the different nanodomains and
offer a very high spatial resolution, as inelastically scattered
waves with different wavelength no longer interfere constructively
or destructively.



\section{Matching of coherent properties\label{matching}}
Clusters, rafts, nanodomains, and patches have become a central
issue in cell membrane studies. The heterogeneous organization of
membrane constituents is believed to be essential for cellular
functions such as signalling, trafficking and adhesion
\cite{Lenne:2009,Apajalahti:2009}. The experimental observation of
those often very short-lived nanoscale structures is extremely
difficult. Experimental techniques must be capable to simultaneously
access the small length and the fast time scales. In particular
inelastic scattering may be capable to provide the spatial
resolution needed to investigate small nanometer structures, as will
be discussed in the following. Inelastic neutron scattering
experiments point to a possible coexistence of short-lived nanoscale
gel and fluid domains in lipid bilayers, which may be responsible
for critical behavior and determine material properties.

\begin{figure}[]
\centering
\resizebox{1.00\columnwidth}{!}{\rotatebox{0}{\includegraphics{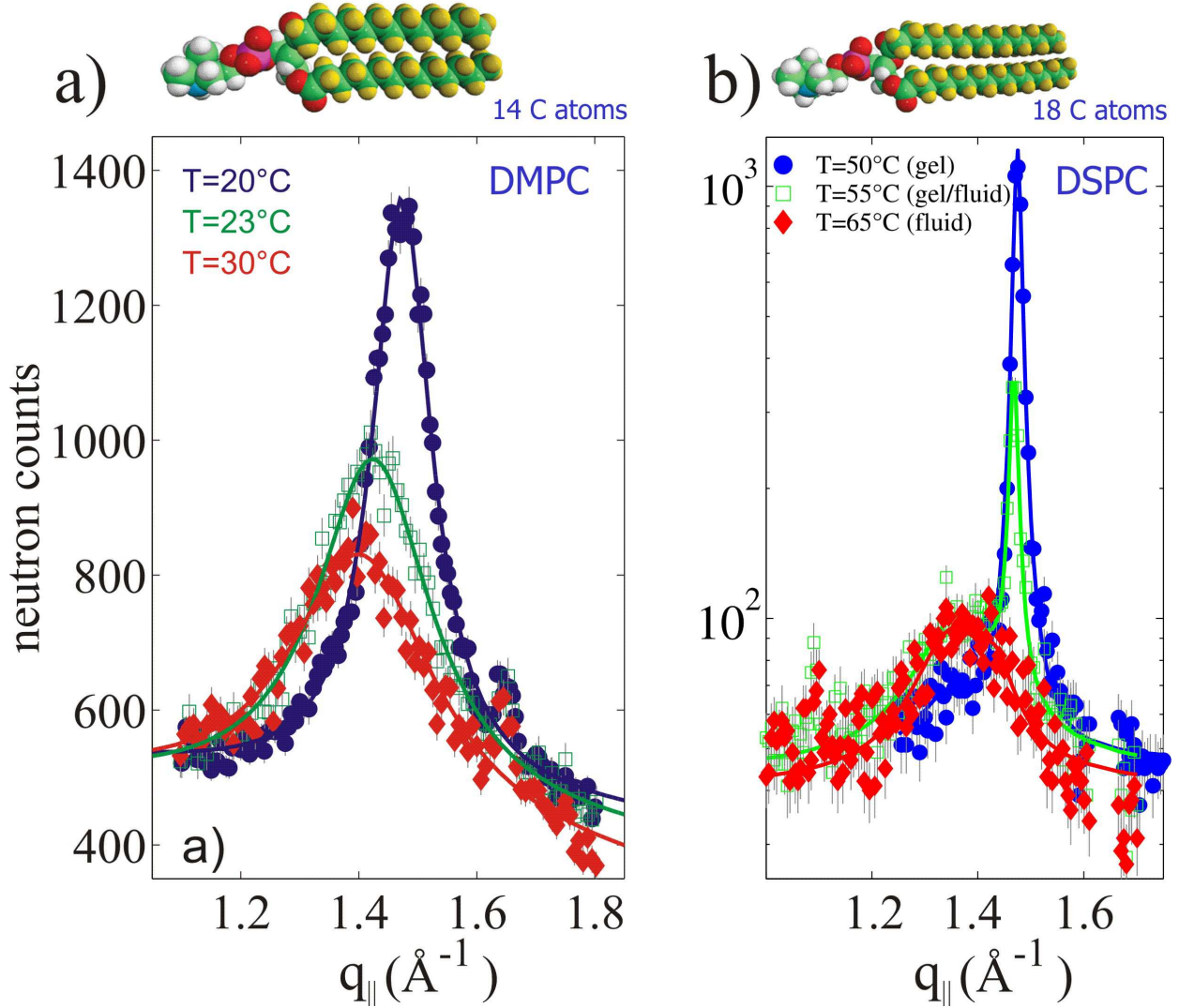}}}
\caption[]{(Color online). (a): Elastic structure factor of the
inter-chain correlation peak in DMPC, measured at temperatures T=20,
23 and 30$^{\circ}$C. (b): Inter-chain correlation peak in
DSPC, at T=50, 55 and 65$^{\circ}$C.} 
\label{kettenpeak}
\end{figure}
\begin{figure}[]
\centering
\resizebox{1.00\columnwidth}{!}{\rotatebox{0}{\includegraphics{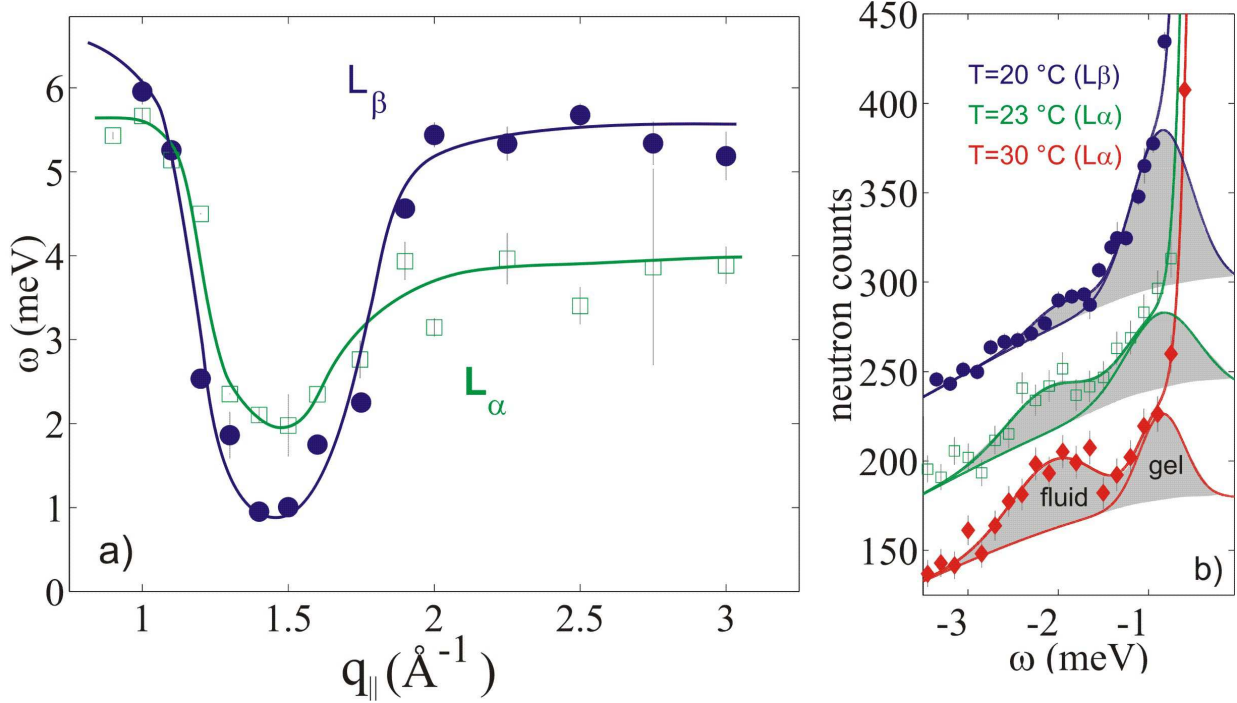}}}
\caption[]{(Color online). (a) Phospholipid dispersion relation in
the gel and the fluid phase of the DMPC bilayer. (b)  Energy scans
in the dispersion minimum at $q_{||}$=1.5 \AA$^{-1}$ for
temperatures T=20, 23 and 30$^{\circ}$C . (from
\cite{RheinstaedterPRL:2004})}\label{dispersion}
\end{figure}
\begin{figure*}[]
\centering
\includegraphics[width=0.75\textwidth,angle=0]{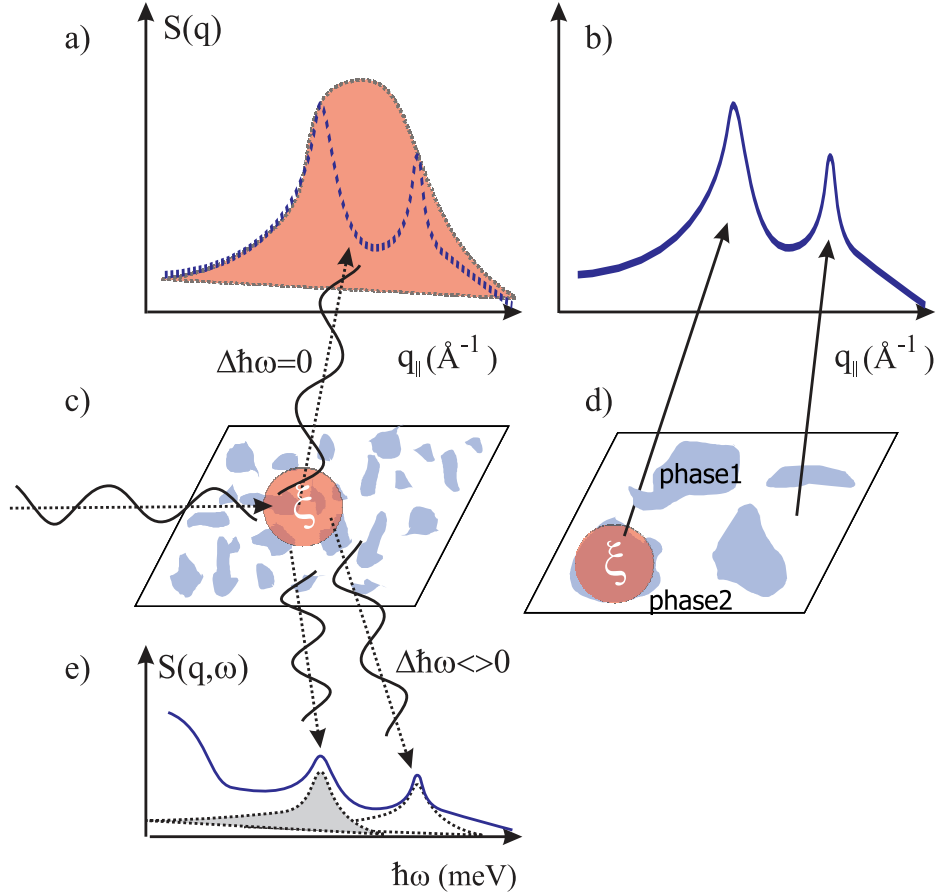}
\caption[]{(Color online). For domain sizes smaller than the
coherence length of the scattering probe (c), diffraction data show
averaged values (a), while excitations coexist (e). Large domains
(d) show coexistence in elastic and inelastic data (b).}
\label{nanodomains}
\end{figure*}
The transition from the rigid gel into the fluid phase in
phospholipid bilayers was reported to show a critical behavior
(``critical swelling'') \cite{Chen:1997}. The lamellar spacing $d_z$
of the membrane stacks, i.e., the distance between neighboring
stacked bilayers (normal to the plane of the membrane), but also the
nearest neighbor distance of lipid molecules in the plane of the
membranes and the corresponding coherence length mimics the critical
behavior known from for instance second order magnetic
order-disorder transitions. 
Inelastic neutron experiments in
1,2-dimyristoyl-sn-glycero-3-phosphocholine (DMPC) showed evidence
for a coexistence of very small nanometer gel and fluid domains in
the range of the main phase transition
\cite{RheinstaedterPRL:2004,RheinstaedterPRL:2006}. Coexisting gel
and fluid excitations were found. The long wavelength dispersion
relation showed showed a soft mode at a length scale comparable to
the expected domain size. It was speculated that these domains are
responsible for the extreme softness, the very low value of the
bending modulus of phospholipid bilayers in the range of the phase
transition, the well known ``critical softening'' \cite{Chu:2005}.
Bending of the membrane would occur mainly in the interface between
two nanodomains, which costs less energy than bending a gel or fluid
domain. A corresponding soft mode was found in the long wavelength
spectrum at a length scale of about 45~nm
\cite{RheinstaedterPRL:2006}. A structural feature was also found in
AFM experiments in thick membrane stacks \cite{Schafer:2008}. The
existence of gel and fluid nanodomains is very difficult to prove
experimentally. The reason is the combination of small nanometer
length scales and the high dynamics in the fluid phase of the
membranes with a rapidly
changing domain pattern. 

To clarify this point, we investigated the packing and correlations
of the lipid acyl chains in phospholipid bilayers with different
chain length using neutron diffraction. DMPC, with 14 C-atoms in the
acyl chains, and 1,2-distearoyl-sn-glycero-3-phosphocholine (DSPC),
with 18 C-atoms, were chosen for this study. The experiment was
carried out at the cold triple-axis spectrometer V2 at the
Hahn-Meitner-Institute, Berlin, Germany. Partially, chain deuterated
lipids (DMPC-d54 and DSPC-d70), hydrated by heavy water (D$_2$O),
were used to enhance the intensity of the the lipid acyl chain
correlation peak. The samples were kept in a closed temperature and
humidity controlled aluminum chamber. Hydration of the lipid
membranes was achieved by separately adjusting two heating baths,
connected to the sample chamber and to a heavy water reservoir,
hydrating the sample from the vapor phase. Temperature and humidity
sensors were installed close to the sample. 
$d$-spacings of 53~\AA\ in DMPC-d54 were achieved. The corresponding
relative humidity (RH) can be determined from $d$--RH curves (for
instance from Fig.~3 in \cite{Chu:2005}), to be 99.5\%. No $d$--RH
curves are published for DSPC but we argue that a similar level of
hydration was achieved for the DSPC sample, too. Note that even that
reasonable hydration levels were achieved, the membranes can not be
considered as fully hydrated, with water being in excess, what will
be important for the discussion of the nanodomains below. The main
transition in DMPC occurs at 23.4$^{\circ}$C
 \cite{Jenskea:2008}. The transition temperature, $T_m$, is slightly
lowered in the deuterated compound (DMPC-d54) to about 21$^{\circ}$C
\cite{Aussenac:2003}. Position and intensity of the inter acyl chain
correlation peak was used in neutron diffraction experiments to
define the temperature of the main transition and $T_m$ for DMPC-d54
was determined to 21.5$^{\circ}$C \cite{RheinstaedterPRL:2004}. The
main transition in DSPC was reported to occur at about 54$^{\circ}$C
\cite{Heimburg:2007}. No lowering of the transition temperature was
found in the deuterated compound DSPC-d70 by following the chain
correlation peak by neutron diffraction, and $T_m$ was determined to
54$^{\circ}$C.

The static structure factor $S(q_{||})$ for DMPC and DSPC is shown
in Figure~\ref{kettenpeak} (a) and (b). Data were taken in the
respective gel and fluid phases, and at an intermediate temperature
slightly above the main transition temperature. DMPC showed the well
known critical behavior, with a continuous shift of the peak
position and continuous increase of the peak width when heating from
the rigid low temperature gel into the fluid phase. For DSPC, there
is a coexistence of gel- and fluid domains indicated by the
existence of two well separated peaks in the range of the phase
transition. The weight of the two peaks changes pointing to an
increasing number of fluid domains when heating from the gel into
the fluid phase. Figure~\ref{dispersion} shows the collective short
wavelength dynamics in DMPC-d54. Figure~\ref{dispersion} (a) depicts
the corresponding dispersion relations in the gel and fluid phase.
Temperature dependent inelastic scans are shown in
Figure~\ref{dispersion} (b). The most striking observation is a
coexistence of gel and fluid excitations (with changing population)
in inelastic experiments, while elastic scattering showed a gradual
change of the correlation peak position.

Taking into account a coherence length of the cold neutrons used for
this study of about $\xi$=500~\AA, the results point to a
coexistence of gel- and fluid domains in the range of the phase
transition, but with distinctly different sizes for the two lipids.
The DMPC domains must be distinctly smaller than $\xi$ to average
over the different nearest neighbor distances. The mismatch of the
two phases (and corresponding energies) seems to be larger for
longer acyl chains and the system tries to avoid domain boundaries
by larger domain sizes ($\gg\xi$) in DSPC. The observations are
summarized in Figure~\ref{nanodomains}: The coexisting gel and fluid
nanodomains in DMPC most likely have sizes smaller than the neutron
coherence length $\xi$ (Figure~\ref{nanodomains} (c)). The elastic
scan does not show two separated gel and fluid peaks, but averages
over gel and fluid areas, as shown in Figure~\ref{nanodomains} (a).
As more and more fluid domains are created, the peak gradually
shifts from the gel to the fluid position. While constructive
interference of elastically scattered neutrons leads to the
correlation peaks in $S(q_{||})$, inelastically scattered neutrons
have slightly different final energies and thus show two separate
excitations, as depicted in Figure~\ref{nanodomains} (e). In DSPC
the coexisting nanodomains can be speculated to be larger than $\xi$
so that the elastic data show two distinct peaks. The weight of the
gel peak decreases indicating the creation of more and more fluid
domains when heating though the main transition
(Figures~\ref{nanodomains} (b) and (d)).
If it comes to soft-matter and biological materials, (highly
coherent) high resolution radiation may average over small scale
structures, such as nanodomains. Inelastically scattered waves do no
longer interfere constructively or destructively so that the
scattering volume is drastically reduced, and inelastic
signals cna provide very local information. 

The conclusion about existence and possible impact of nanodomains
observed in phospholipid model membranes on real biological
membranes critically depends on the hydration level achieved in the
scattering experiments. In these experiments, the bilayers are often
hydrated from the vapor phase because the additional scattering and
absorption of bulk water around the membranes increases background
and decreases sample signals considerably. Even that very reasonable
hydration levels can be achieved (note that even full hydration was
reported \cite{KatsarasCell:2000,Katsaras:1998}), the lipid/water
phase diagram and corresponding phases critically depend on the
level of hydration. Following the Gibb's Phase Rule in condensed
systems (without a coexisting gas phase) the number of independent
intensive properties (F), such as temperature, depends on the number
of components (C), and the number of coexisting phases (P): F=C-P+1.
The observation of coexisting domains over a certain temperature
interval implies F=1. The system is composed of lipids and water
(C=2). The number of possible phases then follows to:
P=C+1-F=2+1-1=2. The observation of two lipid phases would therefore
exclude the existence of a coexisting bulk water phase. Conversely,
a membrane completely immersed in a solvent may not show coexisting
gel and fluid areas. However, the observation of small domains in
AFM experiments under excess water \cite{Schafer:2008} possibly
points to their existence in membranes in liquid environment. But
scattering experiments of phospholipid bilayers immersed in water
will be conducted in the future.

\section{Conclusion}
The coherent properties of photons and neutrons are used to study
structure and dynamics in elastic and inelastic x-ray and neutron
scattering experiments. A good monochromaticity is a prerequisite
for atomic resolution in crystal and protein structure
determinations. High resolution radiation is also highly coherent
with large coherence lengths, possibly of several thousands of
Angstroms. The large coherence length may average over small
structures, such as nanoscale domains. While elastic scattering
experiments show average values for peak positions for instance,
inelastic experiments can still provide very local information. The
well known critical behavior in phospholipid bilayers may stem from
coexisting gel and fluid nanodomains in the range of the gel-fluid
transition. Nanodomains are not only important to understand
fundamental properties of model membranes, but also to better
understand complex biological membranes and for instance the
formation and function of lipid rafts
\cite{Eggeling:2009,Oelke:2008,Risselada:2008}. Additional examples
are the recently emerging nanoferroelectrics \cite{Scott:2007}, and
nanoscale magnetic domains in thin magnetic films
\cite{Shpyrko:2007}. The development of neutron instrumentation,
which allows to control the neutron coherence length, can be
envisioned for the future. Such an instrument would not only be
important for the investigation of biological materials, but in all
systems, where fluctuating nanodomains determine material
properties.

\acknowledgments{It is my pleasure to thank Klaus Habicht,
Helmholtz-Zentrum-Berlin, and Beate Br\"uning, who were involved in
the DSPC data collection.}

\bibliography{membranes_09142009}

\end{document}